\documentclass[12pt]{article}

\usepackage[utf8]{inputenc}
\usepackage{amsmath,amssymb,mathtools}
\usepackage{graphicx}
\usepackage{enumerate}
\usepackage{natbib}
\usepackage{url}
\usepackage{lmodern}
\usepackage{authblk}
\usepackage[normalem]{ulem}

\usepackage{amsthm}
\usepackage{appendix}
\usepackage{listings}
\usepackage{float}
\usepackage{mathrsfs}
\usepackage{longtable}
\IfFileExists{algpseudocode.sty}{
\usepackage{algorithm}
\usepackage{algorithmicx}
\usepackage{algpseudocode}
}{
\newfloat{algorithm}{tbp}{loa}
\floatname{algorithm}{Algorithm}
\newcounter{algorithmicstep}
\newlength{\algorithmicindent}
\setlength{\algorithmicindent}{1.5em}
\newenvironment{algorithmic}[1][]{
  \begin{list}{\arabic{algorithmicstep}.}{
    \usecounter{algorithmicstep}
    \setlength{\leftmargin}{2.5em}
    \setlength{\itemsep}{0.2em}
  }
}{\end{list}}
\newcommand{\State}{\item}
\newcommand{\Statex}{\item[]}
\newcommand{\Require}{\item[\textbf{Input:}]}
\newcommand{\For}[1]{\State \textbf{for} ##1 \textbf{do}}
\newcommand{\EndFor}{\State \textbf{end for}}
\newcommand{\Return}{\textbf{return}~}
\newcommand{\Comment}[1]{\hfill $\triangleright$~##1}
}
\floatstyle{ruled}
\restylefloat{algorithm}
\floatstyle{plain}
\usepackage{booktabs}
\usepackage{bm}
\usepackage{mdframed}
\usepackage{footnote}
\usepackage{caption}
\usepackage{subcaption}
\usepackage{tikz}
\usetikzlibrary{arrows.meta}

\theoremstyle{plain}
\newtheorem{theorem}{Theorem}
\newtheorem{proposition}{Proposition}
\newtheorem{corollary}{Corollary}
\theoremstyle{definition}
\newtheorem{assumption}{Assumption}
\theoremstyle{remark}

\usetikzlibrary{positioning,calc}

\title{\textbf{Structure-based Transfer Learning}}
\author[1]{Ho Yi Alexis Ho\thanks{Email: \texttt{hyhoai@connect.ust.hk}}}
\author[1]{Xinzhou Guo\thanks{Corresponding author. Email: \texttt{xinzhoug@ust.hk}}}
\author[2]{Shuoxun Xu\thanks{Corresponding author. Email: \texttt{shuoxunxu\_ucb@berkeley.edu}}}
\affil[1]{Department of Mathematics, The Hong Kong University of Science and Technology,
Hong Kong SAR, P.R.C.}
\affil[2]{Department of Biostatistics and Epidemiology, University of California, Berkeley,
California, U.S.A.}
\date{}

\begin{document}

\def\spacingset#1{\renewcommand{\baselinestretch}{#1}\small\normalsize}
\spacingset{1}

\maketitle
\vspace{-1.5\baselineskip}

\begin{abstract}
Transfer learning improves estimation in a target study using information from related
sources. Classical transfer learning is typically data-based, requiring access to the source data
or to a model fitted on them. Neither is available in many modern studies, as both are
often proprietary or unreported.
What can be transferred instead is structure: summarized information derived from a source, such as the supports of predictors, which is available and interpretable without access to the source data or model.
The prevalence of the Large Language Model (LLM) provides a rich source of such structure information. However, data-based transfer learning can not well utilize it as accessing the data and model of the LLM is often infeasible.
In this paper, we propose a novel transfer learning framework based on structure. In particular, we focus on linear model and aim to use supports generated by multiple agents, such as LLM, to improve
estimation in the target study. Our procedure maps each support into the target parameter
space, aggregates the results with data-driven weights, and applies a sparse correction
against inaccurate supports. We establish interpretable conditions for retaining the target-only Lasso rate and for
a faster convergence rate, together with minimax optimality under joint coverage.
Simulations and a DNA-methylation application illustrate gains and robustness.
\end{abstract}

\noindent\textit{Keywords:} structure-based transfer learning,
high-dimensional sparse regression, model aggregation, large language models,
AI agents, hallucination
\vfill

\newpage
\spacingset{1.45}

\section{Introduction}\label{sec:intro}

Transfer learning aims to improve the performance of statistical learning in a target study
by incorporating information from related source studies. In biomedical studies, the sample
size of the target study is often limited by cost or
by difficulties in patient recruitment. However, there often exist several source studies
that share similarities with the target study in the conditions and
populations examined \citep{fogel2018factors}. In such cases, transfer learning can help
identify significant biomarkers and confirm effective interventions, among
other tasks, in a more powerful way. For example,
transfer learning has been used to improve the diagnostic accuracy of
colorectal cancer from histopathological images \citep{kassani2022deep}.
Besides biomedical studies, transfer learning is also widely adopted in the broad areas of
data science, such as finance \citep{lanzetta2024transfer}, political science
\citep{terechshenko2020comparison} and engineering \citep{xu2023transfer}.
A Google Scholar search for the phrase \textquotedblleft
transfer learning\textquotedblright{} returns more than seven million records as of September
2026.

To conduct transfer learning, one first needs to
determine what information can be incorporated or transferred from the
source studies. Classical statistical transfer learning typically transfers data by
identifying informative source studies, pooling their data with the target one and fitting
models based on the pooled data \citep{viele2014historical}. Based on
this data-based framework, various methods have been proposed for different scenarios, such
as linear models \citep{li2022transfer}, generalized linear models
\citep{tian2023glmtransfer} and semiparametric models
\citep{he2024representation}; see \cite{zhuang2021comprehensive} for a
literature review. The data-based framework is well justified and widely
adopted in practice, such as biomedical named entity recognition \citep{giorgi2018transfer}
and object detection \citep{saito2019strong}. However, the data-based transfer learning
framework typically requires that (i) the source study data can be
accessed and (ii) the data and the working models considered in the
target and source studies are of the same type, both of which may not
be feasible in practice.

One main motivation for this paper is the rapid development of
large language models
(LLMs) in recent years. Trained on a vast amount of text
via self-supervised learning, an
LLM provides a powerful tool to summarize and incorporate information
from relevant source studies in a natural-language manner
\citep{openai2023gpt4}. However, the source data used to train the LLM
are often proprietary and largely undocumented, making data-based transfer
learning infeasible \citep{bender2021parrots}. Moreover, the data type of
the target study is usually non-text and the working model intended for the target study is
often different from that used to train the LLM: a generative model over
token sequences shares no parameter with the model intended for the target study, so there is
no source parameter to pool with or to shrink toward. Other motivating examples
include meta-analyses where source trial data are often private and
inaccessible and source models were predetermined and may not align with the model intended
for the target study \citep{borenstein2021metaanalysis}.

What can be transferred instead is structure.
Here, structure refers to
the summarized information derived from source studies, which is available and interpretable
without accessing the data or knowing the model used in the source study
\citep{candes2018knockoffs}. The prevalence of LLMs
provides a rich source of structure information, such as the groups of
measurements that previously published studies of a related outcome reported as relevant.
Other examples of structure information include, but are not limited to,
summary statistics in meta-analyses and the variable sets supplied by
curated biological knowledge and domain experts, which already guide variable selection in
genomics \citep{helleputte2009feature,zhang2025llmlassorobustframeworkdomaininformed}.
In the application considered in this paper, for example, Claude Opus 5
\citep{anthropic2026opus5} prompted with the identifiers of the $27{,}578$ CpG positions
measured in the saliva DNA-methylation data of \citet{bocklandt2011epigenetic} returns groups
of positions, each corresponding to one previously published study of age prediction from
methylation; each group carries no observations and no fitted values, so it can be supplied
without disclosing the source data or the source model. Such structures often contain useful
information about the source studies and, in
particular, can serve as a valuable alternative to data when transferring data is infeasible
or inapplicable. Transferring
structure can thus help broaden the scope
of transfer learning, better utilize source information and protect
privacy of the source studies.
What structure does not carry, however, is any evidence of its own
reliability: it arrives without a standard error or a confidence statement, so a suggestion
that is well grounded and a suggestion that is fabricated are indistinguishable on delivery.
Incorporating such information into a procedure that provably benefits from it when it is
informative, and does no harm when it is not, is far from automatic.

Two challenges stand in the way, and each of the two most natural approaches fails at
exactly one of them. The first is to control hallucination bias. The most direct way to use
a suggested structure is to restrict estimation to it, but a suggestion
carries no calibrated reliability, since reliably eliciting confidence from an LLM remains
an open problem \citep{kadavath2022know,shorinwa2024uqsurvey}, and a fluent yet unsupported
suggestion \citep{maynez2020faithfulness,openai2023gpt4} can yield a biased restricted fit
that performs worse than ignoring the external information altogether. A procedure must
therefore control the effect of misspecified transferred structure, retaining the target-only
rate under interpretable conditions even when the suggestions carry no
signal. The second challenge is to improve accuracy when the suggested
structures do carry signal. A source rarely supplies a single structure:
it can be queried more than once, and different queries return different structures. We
call such a source an agent, and, following the terminology of mixture-of-agents approaches
\citep{wang2024moa}, we call the resulting problem multi-agent structure-based transfer
learning. The signal is distributed across different structures
rather than contained in any one of them, and each may both omit
informative components and include uninformative ones. Information must
therefore be combined across structures, none of which is reliable on its
own. Simply merging them into a single structure is not enough, as
it provides no mechanism to weight them by usefulness. It retains every uninformative
component that any structure contributes, and it cannot recover the signal that all of them
omit. Meeting the two challenges therefore requires encoding each discrete
structure as a continuous, target-calibrated quantity, aggregating these
quantities with data-driven weights, and separately recovering the signal that
no structure contains.

In this paper, we study multi-agent structure-based transfer learning.
In particular, we focus on high-dimensional sparse linear regression and
propose a method to improve estimation in the target study with the structures supplied by
the agents.
In this setting, estimation hinges on identifying
a small active set \citep{tibshirani1996lasso,buhlmann2011statistics}, so the relevant structure
is a support, that is, a set of predictors declared active. Each external support acts as an externally generated
prior on that active set.
The method combines the supports using the target data alone
in three stages: a restricted least-squares fit encodes each support as a
continuous quantity calibrated to the target; a penalized regression aggregates these
quantities with data-driven weights; and a sparse residual correction recovers the signal
that no support contains, the step that controls hallucination bias. Every stage uses
off-the-shelf software.
We establish the $\ell_2$ convergence rate of the proposed estimator.
When the supports are informative, the estimator improves substantially on
the target-only Lasso; when they are not, the residual correction retains the target-only
Lasso rate under mild conditions. 
Moreover, we establish a matching minimax rate, whose definition relies on the structure
only, and show that the proposed method achieves the minimax rate.

Our method departs from the data-based transfer learning above by requiring
neither source data nor a shared working model, and it is also related to, but distinct
from, several existing lines of work.
First, our method is related to
knowledge transfer \citep{fang2024bayesian,fang2026multiteacher} as both
draw on trained models to improve a target task. However, knowledge transfer transfers
continuous model outputs, whereas the structures considered here are discrete. Second,
our method is related to LLM-guided feature selection
\citep{helleputte2009feature,zhang2025llmlassorobustframeworkdomaininformed} as both
use external knowledge to guide variable selection. However, LLM-guided
feature selection often focuses on feature selection only and lacks rigorous theoretical
guarantees. In particular, no minimax optimality theory is available to
tell whether a procedure makes the best use of external structure as classical optimality theories are typically given source data or coefficients. Third, our method is related
to privacy-preserving
distributed learning
\citep{cai2022individual,ho2024integrative,kairouz2021federated,liu2021integrative,mcmahan2017fedavg}
as both utilize source information. However, privacy-preserving distributed learning requires
access to the raw data of each site to compute summary statistics and assumes models are
similar across different sites.

The contributions of this paper are threefold. First, we propose a transfer
learning framework based on structure, which requires neither the data nor the working model
of the source studies. Second, we propose a method that encodes each discrete structure as a
continuous quantity calibrated to the target study, so that structures can be combined by
weights estimated from the target data alone. Third, we establish rigorous theoretical
guarantees for the proposed estimator, which existing LLM-guided feature selection methods often
lack, 
and show that the proposed estimator attains a new minimax optimality whose definition does not rely on source coefficients, thus paving a novel way to study optimality when only structure information is available.

The remainder of the paper moves from methodology to theory to empirical evaluation. We
first formalize the problem and develop the three-stage estimator, then establish its
theoretical guarantees, and finally
report simulation studies and a real-data analysis
of the DNA-methylation age data with LLM-proposed CpG supports, which illustrate the
practical gains, before concluding with a discussion of limitations, extensions, and
future directions.

\section{Problem Setup}\label{sec:setup}

We now formalize the structure-based transfer learning problem introduced above, in which the only information carried over from the sources is a collection of candidate supports. On the target site we observe a sample of size $n_0$, an outcome vector $Y^{(0)}\in\mathbb{R}^{n_0}$ and a design matrix $X^{(0)}\in\mathbb{R}^{n_0\times p}$, related through the linear model
\[
Y^{(0)} = X^{(0)}\beta_{\mathrm{true}} + \varepsilon,
\qquad \mathbb{E}[\varepsilon\mid X^{(0)}]=0,
\]
where $\beta_{\mathrm{true}}\in\mathbb{R}^{p}$ is the unknown regression coefficient. We work in the high-dimensional regime in which $p$ may be comparable to or much larger than $n_0$, and we assume $\beta_{\mathrm{true}}$ is sparse, with unknown target support $S_0=\mathrm{supp}(\beta_{\mathrm{true}})\subseteq[p]$ of size $s_0=|S_0|$, writing $[p]=\{1,\dots,p\}$.
For $v\in\mathbb R^p$ and
$S\subseteq[p]$, we write $v_S\in\mathbb R^p$ for the vector with $(v_S)_j=v_j\mathbf 1\{j\in S\}$.
The linear model is the canonical and most studied setting of high-dimensional sparse regression and a useful approximation to many more complex models;
we adopt it throughout. Our aim is to estimate $\beta_{\mathrm{true}}$ more
accurately than is possible from the target data alone, using external information about
which coefficients are nonzero. We now specify the form this information takes.

In addition to the target data, we are given $K$ support sets
$S_1,\dots,S_K\subseteq[p]$, supplied by the agents of
Section~\ref{sec:intro}, and we write
$s_k=|S_k|$ and $s_{\max}=\max_{1\le k\le K}s_k$.
In our real data
application, for instance, the agent is an LLM and each $S_k$ is the set of CpG positions
it extracts from one previously published age-prediction study. Each $S_k$ is a fixed, given input rather
than a quantity estimated from the target data, and it is all we require of an agent: we
do not observe the data behind $S_k$ or the model that produced it, and we assume nothing
about either. Throughout, however, we assume that no external support is
much larger than the target support, that is, $s_{\max}=O(s_0)$; beyond this, how $S_k$
relates to $S_0$ is left unrestricted. It may coincide
with $S_0$, overlap it while omitting or adding variables, or be unrelated to it,
including the extreme case $S_k\cap S_0=\varnothing$; an agent's support may be accurate,
partially useful, or entirely hallucinated, and which of these holds
cannot be determined in advance.

Our goal is to estimate $\beta_{\mathrm{true}}$ under squared $\ell_2$ loss, that is, to
construct an estimator $\widehat\beta$ with $\|\widehat\beta-\beta_{\mathrm{true}}\|_2^2$
small. The information available for this is exactly the target sample
$(Y^{(0)},X^{(0)})$ and the $K$ sets $S_1,\dots,S_K$, and nothing else. This is what
separates the present setting from the data-based framework. In the data-based
framework, each source contributes a sample $(Y^{(k)},X^{(k)})$ together with a
coefficient vector $\beta^{(k)}$ defined in the same model as the target, so the sources
enlarge the effective sample size and the analysis proceeds by measuring the discrepancy
between $\beta^{(k)}$ and $\beta_{\mathrm{true}}$. In the structure-based framework, a source contributes a subset of $[p]$, which
specifies which coordinates may be active but supplies no observations and no
coefficient values. No external observations are therefore available to pool with the
target sample.
Figure~\ref{fig:compare} contrasts the two frameworks: in panel~(a) the sources enter as
additional data samples alongside the target, while in panel~(b) they enter as supports,
and the target sample provides the only observations.

\begin{figure}[H]
    \centering
\begin{tikzpicture}[
  box/.style={rectangle, draw, minimum width=2.2cm, minimum height=0.5cm, thick,
              inner ysep=1pt},
  bigbox/.style={rectangle, draw, minimum width=1.5cm, minimum height=1.5cm, thick},
  solidline/.style={-{Latex[length=2mm,width=1.4mm]}, thick},
  dashline/.style={-{Latex[length=2mm,width=1.4mm]}, thick, dashed}
]

\begin{scope}
    \node[box] (data1) at (0,0) {\footnotesize Source Data 1};
    \node[box, below=0.75cm of data1] (data2) {\footnotesize Source Data 2};
    \node[below=0.2cm of data2] (dots1) {$\vdots$};
    \node[box, below=0.2cm of dots1] (datak) {\footnotesize Source Data $K$};
    \node[box, below=0.75cm of datak] (datatarget) {\footnotesize Target Data};
    \node[bigbox, right=2.5cm of $(data2)!0.5!(datak)$] (central) {Output};
    \draw[solidline] (data1.east)      -- (central.west);
    \draw[solidline] (data2.east)      -- (central.west);
    \draw[solidline] (datak.east)      -- (central.west);
    \draw[solidline] (datatarget.east) -- (central.west);
    \node at (3.1,-5.5) {(a) Data-based framework};
\end{scope}

\begin{scope}[xshift=8cm]
    \node[box] (struct1) at (0,0) {\footnotesize Support 1};
    \node[box, below=0.75cm of struct1] (struct2) {\footnotesize Support 2};
    \node[below=0.2cm of struct2] (dots3) {$\vdots$};
    \node[box, below=0.2cm of dots3] (structk) {\footnotesize Support $K$};
    \node[box, below=0.75cm of structk] (datatargetb) {\footnotesize Target Data};
    \node[bigbox, right=2.5cm of $(struct2)!0.5!(structk)$] (central2) {Output};
    \draw[dashline]  (struct1.east)     -- (central2.west);
    \draw[dashline]  (struct2.east)     -- (central2.west);
    \draw[dashline]  (structk.east)     -- (central2.west);
    \draw[solidline] (datatargetb.east) -- (central2.west);
    \node at (3.1,-5.5) {(b) Structure-based framework};
\end{scope}
\end{tikzpicture}
    \caption{Comparison of the classical data-based transfer learning framework (a) and
    the proposed structure-based framework (b). Solid arrows carry observations; dashed
    arrows carry a support, that is, a subset of $[p]$ with no observations and no
    coefficient values attached. In (b) the target sample provides the only
    observations.}
    \label{fig:compare}
\end{figure}

The unknown reliability of the supports
gives rise to two challenges. Two natural ways of using the supports
illustrate the difficulty of addressing them simultaneously.
The first challenge is
safety. Since no support is known to be reliable, the estimator must not be misled when
the supports are wrong, and must retain the target-only Lasso rate $s_0\log p/n_0$ in the
worst case. Fitting a restricted least-squares regression on a single support $S_k$ fails
to address this challenge: if $S_k$ omits part of $S_0$, the
resulting estimator is biased by an amount that does not vanish as $n_0$ grows. The
second challenge is improvement. When the supports do carry signal,
that signal is distributed across several of them and each is partial, so no support can
be assumed to contain $S_0$ and each may omit active coordinates while including
inactive ones. Fitting on the union $\bigcup_k S_k$
can improve estimation when the union is small and contains $S_0$,
but it does not generally resolve this challenge: it treats the $K$ supports as a
single variable set, with no way to weight them by usefulness; its cardinality can
approach $p$; and any $j\in S_0\setminus\bigcup_k S_k$ remains unrecoverable.
Both challenges arise from the discrete nature of the supports but
generally do not appear when the transferred information is continuous. This motivates
a continuous encoding of each support before aggregation. The estimator in
Section~\ref{sec:method} is designed to address both challenges simultaneously.

\section{Methodology}\label{sec:method}

We now describe the proposed Structure-based Transfer Learning (STL) estimator of $\beta_{\mathrm{true}}$, which exploits the given supports $S_1,\dots,S_K$ without any source-level data, using the target sample $(Y^{(0)},X^{(0)})$ alone.
Our key idea for meeting the two challenges of Section~\ref{sec:setup} is to encode each discrete support, via a local learner, as a continuous target-calibrated predictor, aggregate these predictors with data-driven weights, and add a sparse residual that recovers whatever the supports miss, thereby eliminating hallucination bias.

\begin{algorithm}
\spacingset{1}
\footnotesize
\caption{Structure-based transfer learning (STL) estimator}
\label{alg:TM_SVDLasso_crossfit}
\begin{algorithmic}[1]
\Require target data $(Y^{(0)},X^{(0)})$ with $n_0$ observations; supports $S_1,\dots,S_K$; penalties $\lambda_{\alpha_r},\lambda_{\alpha_{\mathrm{rem}}},\lambda_\delta$; rank threshold $t_{\mathrm{cut}}>0$.
\Statex \textbf{Stage 1: cross-fitted base estimators (local learners)}
\State Split $\{1,\dots,n_0\}$ into two index sets $I_1,I_2$ of approximately equal size; set $I_{-1}=I_2$ and $I_{-2}=I_1$.
\For{$\ell=1,2$}\Comment{cross-fitting}
    \For{$k=1,\dots,K$}
        \State $\displaystyle \widehat\beta^{(-\ell)}_{S_k}\gets\arg\min_{\beta\in\mathbb{R}^p:\,\beta_{S_k^c}=0}\frac{1}{2|I_{-\ell}|}\bigl\lVert Y^{(0)}_{I_{-\ell}}-X^{(0)}_{I_{-\ell}}\beta\bigr\rVert_2^2$ \Comment{local learner on $S_k$}
    \EndFor
    \State Set $\widehat B^{(-\ell)}=\bigl[\widehat\beta^{(-\ell)}_{S_1},\dots,\widehat\beta^{(-\ell)}_{S_K}\bigr]\in\mathbb{R}^{p\times K}$ and $\widehat Z^{(\ell)}=X^{(0)}_{I_\ell}\widehat B^{(-\ell)}\in\mathbb{R}^{|I_\ell|\times K}$.
    \Statex \textbf{Stage 2: SVD-guided aggregation} \textit{(fold $\ell$)}
    \State Compute the SVD $\widehat Z^{(\ell)}=\widehat U^{(\ell)}\widehat\Lambda^{(\ell)}\bigl(\widehat V^{(\ell)}\bigr)^{\top}$ with singular values $\widehat\sigma^{(\ell)}_1\ge\cdots\ge\widehat\sigma^{(\ell)}_K$.
    \State Set the effective rank $\widehat r^{(\ell)}=\operatorname{card}\bigl\{i:\widehat\sigma^{(\ell)}_i/\sqrt{|I_\ell|}>t_{\mathrm{cut}}\bigr\}$.
    \State Let $\widehat V^{(\ell)}_r$ be the first $\widehat r^{(\ell)}$ columns of $\widehat V^{(\ell)}$ and $\widehat V^{(\ell)}_{\mathrm{rem}}$ the remaining $K-\widehat r^{(\ell)}$ columns, and set the projectors $P^{(\ell)}_r=\widehat V^{(\ell)}_r(\widehat V^{(\ell)}_r)^{\top}$ and $P^{(\ell)}_{\mathrm{rem}}=\widehat V^{(\ell)}_{\mathrm{rem}}(\widehat V^{(\ell)}_{\mathrm{rem}})^{\top}$.
    \State Compute
    \Statex \hspace*{\algorithmicindent}$\displaystyle\widehat\alpha^{(\ell)}=\arg\min_{\alpha\in\mathbb{R}^K}\ \frac{1}{2|I_\ell|}\bigl\lVert Y^{(0)}_{I_\ell}-\widehat Z^{(\ell)}\alpha\bigr\rVert_2^2+\lambda_{\alpha_r}\bigl\lVert P^{(\ell)}_r\alpha\bigr\rVert_1+\lambda_{\alpha_{\mathrm{rem}}}\bigl\lVert P^{(\ell)}_{\mathrm{rem}}\alpha\bigr\rVert_1.$
\EndFor
\State $\displaystyle\widehat\beta^{\mathrm{pre}}=\frac{1}{2}\sum_{\ell=1}^{2}\widehat B^{(-\ell)}\widehat\alpha^{(\ell)}$.
\Statex \textbf{Stage 3: sparse residual correction}
\State Compute
\Statex \hspace*{\algorithmicindent}$\displaystyle\widehat\delta=\arg\min_{\delta\in\mathbb{R}^p}\ \frac{1}{2n_0}\bigl\lVert Y^{(0)}-X^{(0)}(\widehat\beta^{\mathrm{pre}}+\delta)\bigr\rVert_2^2+\lambda_\delta\lVert\delta\rVert_1.$
\State \Return $\widehat\beta^{STL}=\widehat\beta^{\mathrm{pre}}+\widehat\delta$.
\end{algorithmic}
\end{algorithm}

The estimator proceeds in three stages. In the first stage, we fit a
least-squares regression on the target data restricted to the covariates in each
$S_k$. This encodes each discrete support as a continuous
quantity lying in the target parameter space, which makes the remaining two stages possible:
the supports can now be weighted rather than merely combined as sets. In the second stage, we
multiply the $K$ fitted coefficient vectors by $X^{(0)}$ to form a synthetic
feature matrix with $K$ columns. We then aggregate
these columns into a preliminary estimator $\widehat\beta^{\mathrm{pre}}$
through a penalized regression whose penalty is guided by the singular value decomposition of
this synthetic feature matrix.
Estimating the weights from the target data addresses the improvement challenge, since a
source is relied upon in proportion to its usefulness and the search is carried out in $K$
dimensions rather than $p$; the singular value decomposition keeps those weights stable when
overlapping supports make the columns collinear. In the third stage, we fit a sparse
correction $\widehat\delta$ to the residual of $\widehat\beta^{\mathrm{pre}}$ on the target
sample and return $\widehat\beta^{STL}=\widehat\beta^{\mathrm{pre}}+\widehat\delta$. This
addresses the safety challenge: it recovers the active covariates that no support contains,
so the estimator is not confined to what the agents propose and is not misled when they are
wrong. The full procedure is summarized in Algorithm~\ref{alg:TM_SVDLasso_crossfit}.

Several distinguishing features and implementation details are worth
noting. First, the restricted least-squares fits act as a continuous encoder of
the discrete supports. Unlike existing approaches that operate on the supports directly
through set operations, such as taking their union or intersection, the encoder allows each
support to be weighted by its usefulness and calibrates it on the target sample, so that the
structure is used both efficiently and correctly.
Second, the singular value
decomposition avoids the nonidentifiability induced by overlapping supports: their synthetic
features can be collinear, so that the aggregation weights are not identified, and penalizing
the principal and remainder directions separately resolves this. This step may be skipped
in practice when the external supports are disjoint, since the
synthetic feature matrix then has full column rank and the penalty reduces to
an ordinary $\ell_1$ penalty.
Third, the weights and the correction are learned by Lasso
fits designed for both statistical and computational efficiency. In particular, the
aggregation needs only a single $K$-dimensional Lasso in each fold, whose statistical cost
grows with $\log K$ rather than $\log p$, and the only $p$-dimensional fit is the residual
correction. Since this correction starts from $\widehat\beta^{\mathrm{pre}}$, the remaining
residual is small when the supports are informative, and the correction converges in few
iterations, so that the procedure can be faster than a Lasso fit on the target data alone.

\section{Theoretical Properties}\label{sec:theory}

We state the assumptions in Section~\ref{sec:assumptions} and establish the
convergence rate of $\widehat\beta^{STL}$ in Section~\ref{sec:rate}.
In Section~\ref{sec:implications}, we derive the implications of this rate for
safety and improvement. Finally, in Section~\ref{sec:optimality}, we show that the
rate is minimax optimal. Proofs are given in the Supplementary Material.

\subsection{Assumptions}\label{sec:assumptions}

We impose three assumptions. Assumption~\ref{ass:subg} is a standard
sub-Gaussian condition on the target design and errors.
Assumptions~\ref{ass:sep} and~\ref{ass:stability} concern the external supports:
Assumption~\ref{ass:sep} requires that the source directions they induce be
distinguishable, and Assumption~\ref{ass:stability} requires that the signal they omit
and the aggregation weights they receive be bounded.
Assumptions~\ref{ass:sep} and~\ref{ass:stability} are needed only for the
improvement, not for the safety guarantee in Corollary~\ref{cor:safety}, and hold in a broad
range of settings.

We first introduce the notation used in these assumptions and in the results that follow.
Let $\Sigma_X^{(0)}$ denote the covariance matrix of the target predictors,
and write $\|v\|_{\Sigma_X^{(0)}}^2=v^\top\Sigma_X^{(0)}v$ for $v\in\mathbb{R}^p$.
For each support $S_k$, let
$\beta^*_{S_k}=\arg\min_{b\in\mathbb{R}^p:\,\mathrm{supp}(b)\subseteq S_k}
\|\beta_{\mathrm{true}}-b\|_{\Sigma_X^{(0)}}^2$
be its population restricted regression coefficient.
Define the matrix of source directions
$B^*=[\beta^*_{S_1},\ldots,\beta^*_{S_K}]\in\mathbb{R}^{p\times K}$ and
$r=\operatorname{rank}\{(\Sigma_X^{(0)})^{1/2}B^*\}$.
Let $\alpha^*_{B^*}$ be the minimum-$\ell_2$-norm minimizer of
$\|\beta_{\mathrm{true}}-B^*\alpha\|_{\Sigma_X^{(0)}}^2$
over $\alpha\in\mathbb{R}^K$, and set
$s_r=\|\alpha^*_{B^*}\|_0$ and
$\Delta=\|\beta_{\mathrm{true}}-B^*\alpha^*_{B^*}\|_2$.
Thus $\Delta$ is the distance to the best population source-based approximation,
$s_r$ is the number of nonzero population aggregation weights, and $r$ is the
number of linearly independent source directions. The minimum-norm choice identifies the weights
when repeated supports make them non-unique. With this notation, we
now state the assumptions.

\begin{samepage}
\begin{assumption}[Sub-Gaussian design and noise]\label{ass:subg}
For fixed constants $0<\kappa_-\le\kappa_+<\infty$ and $L,L'<\infty$,
the design and errors satisfy the following:
(i) the rows $X_i^{(0)}$ are i.i.d.\ centered sub-Gaussian vectors with covariance
$\Sigma_X^{(0)}$, where $\kappa_-I_p\preceq\Sigma_X^{(0)}\preceq\kappa_+I_p$ and
$\|v^\top X_i^{(0)}\|_{\psi_2}\le L\|v\|_{\Sigma_X^{(0)}}$ for all $v\in\mathbb{R}^p$; and
(ii) the errors $\varepsilon_i$ are i.i.d.\ centered sub-Gaussian variables,
independent of $X^{(0)}$, with fixed variance $\sigma^2>0$ and
$\|\varepsilon_i\|_{\psi_2}\le L'$.
\end{assumption}
\end{samepage}

\begin{samepage}
\begin{assumption}[Distinguishable source directions]\label{ass:sep}
Let $B^*_{\mathrm d}\in\mathbb R^{p\times|\{S_1,\ldots,S_K\}|}$ be the matrix obtained
from $B^*$ by keeping only one column for each repeated support,
and let $\sigma_{\min}\{\cdot\}$ and $\sigma_{\max}\{\cdot\}$ denote the smallest and
largest singular values of a matrix.
For
fixed constants $c_{\mathrm{sig}}>0$ and $c'_{\mathrm{sig}}<\infty$,
\[
\text{(i)}\ \sigma_{\min}\{(\Sigma_X^{(0)})^{1/2}B^*_{\mathrm d}\}\ge c_{\mathrm{sig}};
\quad\text{and}\quad
\text{(ii)}\ \frac{\sigma_{\max}\{(\Sigma_X^{(0)})^{1/2}B^*_{\mathrm d}\}}
{\sigma_{\min}\{(\Sigma_X^{(0)})^{1/2}B^*_{\mathrm d}\}}\le c'_{\mathrm{sig}}.
\]
\end{assumption}
\end{samepage}

\begin{samepage}
\begin{assumption}[Bounded omitted signal and aggregation weights]\label{ass:stability}
For a fixed constant $0<M<\infty$, the external supports satisfy the following:
\[
\text{(i)}\ \min_{k\in[K]}\|\beta_{\mathrm{true}}-\beta^*_{S_k}\|_{\Sigma_X^{(0)}}
\le M;
\quad\text{and}\quad
\text{(ii)}\ \max_{k\in[K]}\bigl(\|\alpha^*_{B^*}\|_1^2\,
\|\beta_{\mathrm{true}}-\beta^*_{S_k}\|_{\Sigma_X^{(0)}}^2\bigr)
\le M\frac{K}{r}.
\]
\end{assumption}
\end{samepage}

Assumption~\ref{ass:subg} requires sub-Gaussian tails for the predictors
and the errors, and bounded eigenvalues for the predictor covariance. These conditions are
standard in high-dimensional regression
\citep{buhlmann2011statistics,vershynin2018high,wainwright2019high}. They hold for a broad
class of designs, including Gaussian designs, predictor vectors with independent bounded
coordinates, and nonsingular linear transformations of either.

Assumption~\ref{ass:sep} imposes two requirements on the source directions
of the distinct supports, that is, after repeated supports are removed.
Assumption~\ref{ass:sep} (i) requires each distinct support to capture sufficient
signal of its own.
This is an eigenvalue condition of the type used in Lasso theory
\citep{bickel2009simultaneous,buhlmann2011statistics}, imposed here on
$(\Sigma_X^{(0)})^{1/2}B^*_{\mathrm d}$, which has at most $K$ columns,
rather than on the $p$ columns of $X^{(0)}$.
Assumption~\ref{ass:sep} (ii)
bounds the condition number of $(\Sigma_X^{(0)})^{1/2}B^*_{\mathrm d}$,
so that the distinct source directions are of comparable strength and not nearly
collinear. Bounded condition numbers are standard in high-dimensional regression
\citep{buhlmann2011statistics,wainwright2019high}. 
The condition is also mild: it is
invariant to the scale of $\beta_{\mathrm{true}}$ and
constrains only how the distinct source directions relate to one another, not how
accurate the supports are.
Taken together, the two requirements in Assumption \ref{ass:sep}
hold in a broad range of settings allowing repeated supports, as long as each distinct support captures enough and separable signal of its own.
In particular, when $\Sigma_X^{(0)}=I_p$, Assumption~\ref{ass:sep} is satisfied either (a)
when the distinct supports are disjoint and the active part of each support carries
sufficient signal, or (b) when the distinct supports overlap, but the part of each
support not shared with the others carries sufficient signal, at least of the
same signal order as that on the shared part.

Assumption~\ref{ass:stability} controls the signal omitted by the supports
together with the aggregation weights.
Assumption~\ref{ass:stability} (i) bounds the signal omitted by the regression
restricted to the most accurate support.
This condition is mild, as it requires only one support to
be sufficiently reliable.
Assumption~\ref{ass:stability} (ii)
bounds the largest signal omitted by any support, scaled by the total magnitude of the
population aggregation weights.
This condition restricts only the product of the largest
omitted signal and the total weight, so the two can trade off against each other to some
extent.
Both conditions hold, for example, when the signal-to-noise ratio is bounded
and either a few supports carry most of the signal or many supports overlap
substantially, so that the aggregate needs only a small weight on each of them.

\subsection{Convergence rate} \label{sec:rate}

We now establish the convergence rate of $\widehat\beta^{STL}$. The rate
has three terms, corresponding to the three stages of
Algorithm~\ref{alg:TM_SVDLasso_crossfit}: a projection term for the
restricted regressions on
the target sample, an aggregation term for combining the supports, and a bias term governed
by the approximation error $\Delta$. In Section~\ref{sec:implications}, we use this rate to derive the safety and
improvement properties, and we give an interpretable bound on $\Delta$. In
Section~\ref{sec:optimality}, we show that the convergence rate of $\widehat\beta^{STL}$ is
minimax optimal.

Take fixed, sufficiently large constants $A_r,A_\delta$ in
\begin{equation}\label{eq:tuning}
\begin{gathered}
\lambda_{\alpha_r}=A_r\sigma
\max\!\left\{\sigma,\max_{k\in[K]}\|\beta^*_{S_k}\|_{\Sigma_X^{(0)}}\right\}
\sqrt{\frac{\log(2K)}{n_0}},
\qquad \lambda_{\alpha_{\mathrm{rem}}}>0,\\
\lambda_\delta=A_\delta\sigma\sqrt{\frac{\log(2p)}{n_0}},
\qquad 0<t_{\mathrm{cut}}<\frac{c_{\mathrm{sig}}}{4}.
\end{gathered}
\end{equation}
Under Assumption~\ref{ass:sep}, any rank threshold in the range given by
\eqref{eq:tuning} separates the retained source directions from the discarded ones.
The constants in Assumptions~\ref{ass:subg}--\ref{ass:stability} and in
\eqref{eq:tuning} are fixed along the asymptotic sequence.

\begin{samepage}
\begin{theorem}[Convergence rate of $\widehat\beta^{STL}$]\label{thm:main-rate}
Under Assumptions~\ref{ass:subg}--\ref{ass:stability} and the tuning
\eqref{eq:tuning}, along sequences for which $p\to\infty$,
$\sum_{k=1}^K s_k+K=o(n_0)$, and $n_0\ge Cs_0\log(2p)$ for a sufficiently
large fixed $C$,
\begin{equation}\label{eq:main-rate}
\begin{aligned}
\|\widehat\beta^{STL}-\beta_{\mathrm{true}}\|_2^2
=O_{\mathbb P}\bigg(&
\underbrace{\frac{\sigma^2Ks_0}{n_0r}}_{\textnormal{projection}}
+\underbrace{\frac{\sigma^2s_r\log(2K)}{n_0}}_{\textnormal{aggregation}}\\
&+\underbrace{\frac{\sigma^2\|\beta_{\mathrm{true}}-B^*\alpha^*_{B^*}\|_0\log(2p)}{n_0}
\wedge\Delta^2}_{\textnormal{bias}}\bigg).
\end{aligned}
\end{equation}

\end{theorem}
\end{samepage}

The three terms are the cost of fitting the restricted least squares on
the target sample, the cost of aggregating the supports, and the bias left by the supports
that no aggregation can remove.
The projection term is the cost of the $K$ restricted ordinary least
squares fits on the target sample, each of order $s_0/n_0$, divided by
$r$, which under Assumption~\ref{ass:sep} equals the number of distinct
supports.
The factor $K/r$ comes from the bound in
Assumption~\ref{ass:stability} (ii), which controls how the estimation
errors of the restricted regressions accumulate in the aggregate. Under
Assumption~\ref{ass:sep}, $r<K$ only when some supports are repeated, which may make the bound
loose. To tighten the bound in this case, we show in Proposition~S2 of the Supplementary
Material that the factor $K/r$ can be removed under mild additional conditions. When
$r\asymp K$, the two cancel and the term is of order $s_0/n_0$, smaller than the target-only
Lasso rate by a factor of $\log p$, so it never affects the comparison with the Lasso.
The aggregation
term is the cost of selecting and weighting the $K$ sources; because it is paid in the
$K$-dimensional source space rather than in $\mathbb{R}^p$, it depends on the ambient
dimension only through $\log K$, and is the origin of the improvement.
The bias term is where the quality of the
supports enters. Its first part is of the order of the target-only Lasso rate
when $\|\beta_{\mathrm{true}}-B^*\alpha^*_{B^*}\|_0=O(s_0)$, for example when
$|(\bigcup_kS_k)\setminus S_0|=O(s_0)$. Its second part, $\Delta^2$, is small when the
supports approximate $\beta_{\mathrm{true}}$ well.
The result differs from those of data-based transfer learning in what gives the rate
improvement. In data-based transfer, the improvement comes from similarity between the
source and target coefficient vectors under the same model, measured by a contrast such as
$\|\beta^{(k)}-\beta_{\mathrm{true}}\|_1$ \citep{li2022transfer,tian2023glmtransfer}. In our
rate above, the improvement comes from the quality of the supports alone, measured by
$\Delta$. No model is assumed on the source side, and the agents that produced the supports
may be of any kind.

\subsection{Implications: Safety and Improvement}\label{sec:implications}

Theorem~\ref{thm:main-rate} has two implications for the proposed estimator $\widehat\beta^{STL}$, safety and accuracy improvement, corresponding to
the two challenges of Section~\ref{sec:setup}. First, the estimator is
safe: whatever the value of $\Delta$, the residual correction retains the target-only
Lasso rate $s_0\log p/n_0$ under mild conditions, so that hallucinated or irrelevant
supports do not degrade the rate asymptotically. Second, the estimator
attains a faster rate when $\Delta$ is small, as is the case in a broad
range of scenarios. In particular, when $\Delta=0$, the rate reduces to
$Ks_0/(n_0r)+s_r\log K/n_0$, replacing the $\log p$ of the Lasso by $\log K$.
We establish safety first and then bound $\Delta$ to identify when the
improvement occurs.

For safety, Theorem~\ref{thm:main-rate} implies that
$\widehat\beta^{STL}$ retains the target-only Lasso rate regardless of the value of
$\Delta$ when $K$ is fixed.
Corollary~\ref{cor:safety} relaxes these conditions: it establishes safety
for any $K$ under condition~\eqref{eq:safety-condition}. Unlike
Assumptions~\ref{ass:sep} and~\ref{ass:stability}, this condition does not require the external
supports to be informative: it only limits the number of inactive variables they add, so all of
them may miss the entire target support.

\begin{samepage}
\begin{corollary}[Safety under inaccurate supports]\label{cor:safety}
Under Assumption~\ref{ass:subg}, suppose $s_0\ge1$, $p\to\infty$, and
\begin{equation}\label{eq:safety-condition}
\left|\left(\bigcup_{k=1}^K S_k\right)\setminus S_0\right|\le C_0s_0
\end{equation}
for a fixed $C_0$. Suppose further that $n_0\ge C_1 s_0\log(2p)$ and
$\lambda_\delta=A_\delta\sigma\sqrt{\log(2p)/n_0}$, where $C_1$ and $A_\delta$ are
sufficiently large fixed constants and $C_1$ may depend on $C_0$. Then, for any positive
aggregation penalties $\lambda_{\alpha_r},\lambda_{\alpha_{\mathrm{rem}}}$ and any
positive rank threshold $t_{\mathrm{cut}}$,
\begin{equation}\label{eq:safety-rate}
\|\widehat\beta^{STL}-\beta_{\mathrm{true}}\|_2^2
=O_{\mathbb P}\!\left\{\frac{\sigma^2s_0\log(2p)}{n_0}\right\}.
\end{equation}
\end{corollary}
\end{samepage}

We next bound $\Delta$ in Proposition~\ref{prop:Delta-coverage}. For
correlated predictors, the bound involves
\[
\rho_{\mathrm{off}}=
\max_{k\in[K]}\ \max_{i\in S_k,\ j\in S_0\setminus S_k}
|\operatorname{Corr}(X_i,X_j)|,
\]
the largest correlation between a predictor that an external support retains and an active
predictor it omits, with an empty maximum defined as zero.

\begin{samepage}
\begin{proposition}[Bounding the approximation error]\label{prop:Delta-coverage}
Under the covariance bounds in Assumption~\ref{ass:subg},
\begin{equation}\label{eq:single-coverage}
\Delta\le C\min_{k\in[K]}\|(\beta_{\mathrm{true}})_{S_0\setminus S_k}\|_2.
\end{equation}
Suppose, in addition, that $S_0\cap\bigcup_{k=1}^K S_k$ is nonempty and that, for a fixed
$c\in(0,1]$,
\begin{equation}\label{eq:balanced-coverage}
\sum_{k=1}^K\mathbf1\{j\in S_k\}
=\sum_{k=1}^K\mathbf1\{j'\in S_k\}\ge cK,
\qquad \text{for all } j,j'\in S_0\cap\bigcup_{k=1}^K S_k.
\end{equation}
Then
\begin{equation}\label{eq:coverage-bound}
\Delta\le C\left\{
\|(\beta_{\mathrm{true}})_{S_0\setminus\bigcup_{k=1}^K S_k}\|_2
+\frac1K\sum_{k=1}^K
\min\{1,\rho_{\mathrm{off}}\sqrt{s_k|S_0\setminus S_k|}\}
\|(\beta_{\mathrm{true}})_{S_0\setminus S_k}\|_2\right\},
\end{equation}
where $C$ depends only on the covariance bounds and on $c$. If
$\Sigma_X^{(0)}=I_p$, then $\rho_{\mathrm{off}}=0$, so the second term in
\eqref{eq:coverage-bound} vanishes and the bound holds with equality for $C=1$:
\begin{equation}\label{eq:balanced-delta}
\Delta^2=\sum_{j\in S_0\setminus\bigcup_{k=1}^K S_k}(\beta_{\mathrm{true}})_j^2.
\end{equation}
\end{proposition}
\end{samepage}

For improvement, Theorem~\ref{thm:main-rate} implies that the
proposed method attains a faster rate as long as $\Delta$ is small and $K$ is not too
large.

Proposition~\ref{prop:Delta-coverage} shows that $\Delta$ is small as
long as the supports, individually or jointly, cover most of the active signal. Specifically,
by \eqref{eq:single-coverage}, $\Delta$ is small when a single support misses little of the
active signal. By \eqref{eq:coverage-bound}, $\Delta$ is also small when the supports jointly
cover most of the active signal, even if none does so on its own, provided that each covered
active coordinate is proposed by the same number of agents and by a nonvanishing proportion of
them, and that the coordinates a support omits are weakly correlated with those it retains. In
particular, $\Delta=0$ whenever $\beta_{\mathrm{true}}$ is a linear combination of the source
directions $\beta^*_{S_k}$. For example, this holds when one support contains $S_0$, since the
bound \eqref{eq:single-coverage} is then zero. If $\Sigma_X^{(0)}=I_p$,
\eqref{eq:balanced-delta} shows that $\Delta=0$ also when the supports jointly cover $S_0$ under
\eqref{eq:balanced-coverage}, although no individual support is complete.

\subsection{Minimax optimality}\label{sec:optimality}

We next show that the convergence rate of the proposed estimator  $\widehat\beta^{STL}$ derived in Theorem~\ref{thm:main-rate} is minimax optimal
over a general parameter class. 
Specifically, we consider the parameter class for each $\bar\Delta\ge0$,
\begin{equation}\label{eq:minimax-class}
\Theta(s_0,\bar\Delta)=\left\{\beta\in\mathbb R^p:
\begin{array}{l}
\|\beta\|_0\le s_0,\quad
\|\beta_{(\bigcup_{k=1}^K S_k)^c}\|_2\le\bar\Delta,\\[3pt]
\beta\text{ satisfies Assumptions~\ref{ass:sep} and~\ref{ass:stability}}
\end{array}\right\}.
\end{equation}
The parameter class is interpreted as follows:
given external supports $S_1,\ldots,S_K$, $\Theta(s_0,\bar\Delta)$ consists
of all $\beta$ that satisfy Assumptions~\ref{ass:sep} and~\ref{ass:stability} together
with two constraints:
(1) $\beta$ is $s_0$-sparse; and (2) $\beta$ has at most $\bar\Delta$ of its signal
outside the union of the external supports.
Here, $\bar\Delta$ can be viewed as an upper bound on the approximation
error $\Delta$ in Theorem~\ref{thm:main-rate}, because, under the conditions of
Theorem~\ref{thm:minimax-rate} below, if $\beta_{\mathrm{true}}\in\Theta(s_0,\bar\Delta)$,
then $\Delta=\|(\beta_{\mathrm{true}})_{(\bigcup_kS_k)^c}\|_2\le\bar\Delta$.

Unlike the class used in data-based transfer learning by \citet{li2022transfer},
$\Theta(s_0,\bar\Delta)$ constrains the $\ell_0$ ball $\{\beta:\|\beta\|_0\le s_0\}$ through
the supports rather than through source coefficients.
In \citet{li2022transfer}, the level $h$
bounds $\|\beta^{(k)}-\beta\|_1$ for the coefficient vector $\beta^{(k)}$ of each informative
source. Such a level presupposes that the sources follow the same linear model as the target and
have coefficient vectors $\beta^{(k)}$. In $\Theta(s_0,\bar\Delta)$, the level $\bar\Delta$
bounds $\|\beta_{(\bigcup_kS_k)^c}\|_2$, which
involves no source coefficient $\beta^{(k)}$, only $\bigcup_kS_k$, and is thus
more practical.

\begin{samepage}
\begin{theorem}[Minimax rate under joint coverage]\label{thm:minimax-rate}
Suppose the external supports are distinct, the rows of $X^{(0)}$ are
independent $N(0,I_p)$, and the entries of $\varepsilon$
are independent $N(0,\sigma^2)$, independently of the design. Under the regularity
conditions in Section~S4.1 of the
Supplementary Material,
\begin{equation}\label{eq:minimax-three-term-lower}
\inf_{\widetilde\beta}\sup_{\beta\in\Theta(s_0,\bar\Delta)}
\mathbb E_\beta\|\widetilde\beta-\beta\|_2^2
\asymp
\frac{\sigma^2K s_0}{n_0r}
+\frac{\sigma^2s_r\log(2K)}{n_0}
+\left(\frac{\sigma^2s_0\log(2p)}{n_0}\wedge\bar\Delta^2\right),
\end{equation}
where the
infimum is over all estimators based on the target sample and the external supports.
\end{theorem}
\end{samepage}

Theorem~\ref{thm:minimax-rate} identifies the minimax rate on this class and, together
with Theorem~\ref{thm:main-rate}, shows that STL attains it in probability.
Recall that $\bar\Delta$ is an upper bound on $\Delta$ for every
$\beta_{\mathrm{true}}\in\Theta(s_0,\bar\Delta)$, so the rate in
Theorem~\ref{thm:main-rate} is at most the minimax rate in Theorem~\ref{thm:minimax-rate}.
Thus no estimator based on the target sample and the external supports can further improve
on STL uniformly over this class beyond constant factors.

In addition to relying only on structure information, another difference between STL and classical minimax-optimal estimators is
adaptivity: STL attains the minimax rate without knowing $\bar\Delta$, whereas classical minimax-optimal estimators
may need to know the parameter that defines the class in advance. Here $\bar\Delta$ measures how much
of the signal in $\beta$ the external supports miss and is often unknown in practice. With a single
tuning that does not depend on $\bar\Delta$, 
STL attains this rate over $\Theta(s_0,\bar\Delta)$ for every $\bar\Delta\ge0$.
In this sense, STL is
adaptively minimax optimal in probability: its rate adjusts automatically to the quality of
the external supports. By contrast, in data-based transfer learning,
\citet{li2022transfer} establish the minimax rate over a class with a given level $h$, and
their oracle estimator must know in advance the set of
sources with $\|\beta^{(k)}-\beta\|_1\le h$, which depends on this fixed $h$ and is typically
unavailable in practice.

\section{Simulations}
\label{sec:simulations}

We evaluate the performance of the proposed STL estimator
(Algorithm~\ref{alg:TM_SVDLasso_crossfit}) through
simulations.
The experiments address two questions: how STL performs under varying
quality of external support, ranging from highly informative to completely
uninformative; and
how performance scales with the number of supports $K$.

\subsection{Settings}\label{sec:sim-settings}

We set the ambient dimension to $p = 300$, target sparsity
$s_0 = 10$, and target sample size $n_0 = 200$.  The true support is
fixed at $S_0 = \{1, 2, \ldots, s_0\}$, and the nonzero coefficients
are drawn once as
$(\beta_{\mathrm{true}})_j \stackrel{\mathrm{iid}}{\sim} N(0,1)$ for
$j \in S_0$ and held fixed across all replications.  In each of the
$N_{\mathrm{rep}} = 100$ replications, the target design matrix
$X^{(0)} \in \mathbb{R}^{n_0 \times p}$ is drawn with i.i.d.\ $N(0,1)$ entries, together
with an independent noise vector
$\varepsilon \sim N(0, I_{n_0})$, producing the response
$Y^{(0)} = X^{(0)} \beta_{\mathrm{true}} + \varepsilon$.  


We are given $K$ candidate support sets
$S_1, \ldots, S_K \subseteq \{1, \ldots, p\}$, each of size $s_0$.
These are constructed by perturbing the true support $S_0$: for a
prescribed overlap fraction $\omega_k \in [0,1]$, the set $S_k$
retains $\lceil \omega_k \cdot s_0 \rceil$ indices from $S_0$ and
replaces the remainder with indices drawn uniformly from $S_0^c$.
When $\omega_k = 1$ the support coincides with $S_0$; when
$\omega_k = 0$ the support is entirely disjoint. All support sets are
fixed prior to the replication loop and are the only external input to
the proposed method; no auxiliary datasets or coefficient estimates are
used.  This setup directly models the practical scenario in which an
LLM proposes candidate variable sets of varying quality: some may
closely match the true active set, while others may be partially or
entirely hallucinated.

The first experiment fixes $K=5$ and considers four scenarios, A to D,
that progressively degrade the support quality: (A) favorable,
$\boldsymbol{\omega}=(0.9,0.9,0.9,1.0,0.0)$, with three supports overlapping $S_0$ by
$90\%$, one identical to $S_0$ and one disjoint from it;
(B) joint coverage,
$\boldsymbol{\omega}=(0.8,0.8,0.8,0.8,0.8)$, in which no support contains $S_0$, but the
supports jointly cover it: each index of $S_0$ appears in only four of the five supports and
is missing from the remaining one;
(C) sparse informative,
$\boldsymbol{\omega}=(0.9,0.0,0.0,0.0,0.0)$, with one support overlapping by $90\%$ and
four disjoint; and (D) uninformative, $\boldsymbol{\omega}=(0.0,0.0,0.0,0.0,0.0)$, with
all five supports disjoint from $S_0$.  
The scenarios thus range from a
single complete support (A) and joint coverage without a complete support (B), both of
which give $\Delta=0$ by Proposition~\ref{prop:Delta-coverage}, to supports that miss part
of the signal (C) or all of it (D).

The second experiment instead varies the number of supports, increasing $K$
from $3$ to $15$ while holding the proportion of informative supports roughly
constant at about $80\%$: for each $K$, one support is identical to $S_0$
($\omega = 1$), about $80\%$ of the remaining supports have $90\%$ overlap
($\omega = 0.9$), and the rest are completely disjoint ($\omega = 0$).  The full
overlap vectors are listed in Table~\ref{tab:sim-varyK}.  This isolates the effect
of growing $K$ from that of changing support quality. All other quantities are as above.

We compare two estimators on the target data $(X^{(0)},Y^{(0)})$: the
proposed STL (Algorithm~\ref{alg:TM_SVDLasso_crossfit}), with tuning parameters
$(\lambda_{\alpha_r},\lambda_{\alpha_{\mathrm{rem}}},\lambda_\delta)$ selected jointly by
three-fold cross-validation over a fixed grid, and the Target Lasso, the standard Lasso
fitted to $(X^{(0)},Y^{(0)})$ alone, with its penalty selected by three-fold
cross-validation and all external support information ignored.

\subsection{Results}\label{sec:sim-results}

Table~\ref{tab:sim-main} reports the $\ell_2$ estimation error for each
method across the four scenarios of the first experiment.
The Target Lasso does not use the supports, so its entries
coincide across scenarios by construction. The error of STL increases as the quality of the supports
degrades from Scenario~A to Scenario~D.

When one of the supports coincides with $S_0$
(Scenario~A), STL
reduces the $\ell_2$ error by about $52\%$ relative to the Target Lasso.
The aggregation places almost
all of its weight on the exact support, so the additional partially overlapping supports
add no information.
When no support contains $S_0$ but the supports cover it jointly
(Scenario~B), STL reduces the error by about $47\%$ relative to
the Target Lasso, nearly as much as in Scenario~A.
When the supports miss part of the signal (Scenario~C), STL
still reduces the error substantially, by about $16\%$
relative to the Target Lasso.
When
they miss all of the signal (Scenario~D), the aggregation penalty
selected by cross-validation sets all aggregation weights to zero in almost all replications,
and the residual correction $\widehat{\delta}$ absorbs the full signal, so that the mean errors
of STL and the Target Lasso coincide to four decimal places, consistent with
the safety guarantee of Corollary~\ref{cor:safety}.

\begin{table}[H]
\centering
\small
\caption{Mean (standard error) of $\ell_2$ estimation error
$\|\widehat\beta - \beta_{\mathrm{true}}\|_2$ over
$N_{\mathrm{rep}} = 100$ replications with $p = 300$, $s_0 = 10$,
$n_0 = 200$, and $K = 5$. Standard errors are computed as
$\mathrm{SD}/\sqrt{N_{\mathrm{rep}}}$. The best result in each scenario is
bolded.}
\label{tab:sim-main}
\begin{tabular}{lcccc}
\toprule
 & \multicolumn{4}{c}{Scenario} \\
\cmidrule(lr){2-5}
Method & A: Favorable & B: Joint coverage & C: Sparse inf.\ & D: Uninformative \\
\midrule
STL
  & \textbf{0.2628 (0.0050)}
  & \textbf{0.2918 (0.0063)}
  & \textbf{0.4584 (0.0089)}
  & \textbf{0.5475 (0.0074)} \\
Target Lasso
  & 0.5475 (0.0074)
  & 0.5475 (0.0074)
  & 0.5475 (0.0074)
  & \textbf{0.5475 (0.0074)} \\
\bottomrule
\end{tabular}
\medskip
\end{table}

Table~\ref{tab:sim-varyK} reports the results of the second experiment, in which $K$ varies.
The Target Lasso is flat at
$0.547$ across all $K$, as expected.  STL increases only mildly
from $0.251$ at $K = 3$
to $0.287$ at $K = 15$, a roughly $14\%$
rise while $K$
quintuples, and remains well below the Lasso throughout.  The modest
increase is consistent with the $\log K$ cost in the aggregation error
of Theorem~\ref{thm:main-rate}: the aggregation is a
$K$-dimensional problem, and the highly overlapping supports produce nearly
collinear columns in $\widehat{Z}^{(\ell)}$, so that most of their variation lies
in a few directions and the additional supports add little to the search.
These results confirm that the method can absorb a growing
number of supports without substantial performance degradation.
\begin{table}[H]
\centering
\caption{Mean (standard error) of $\ell_2$ estimation error as the number of
external supports $K$ varies, with $p = 300$, $s_0 = 10$,
$n_0 = 200$, and $N_{\mathrm{rep}} = 100$ replications; standard errors are
$\mathrm{SD}/\sqrt{N_{\mathrm{rep}}}$. Each overlap
vector lists the values $\omega_1,\dots,\omega_K$ used to construct the
$K$ supports.}
\label{tab:sim-varyK}
\begin{tabular}{clcc}
\toprule
 & & \multicolumn{2}{c}{Method} \\
\cmidrule(lr){3-4}
$K$ & Overlap structure $(\omega_1,\dots,\omega_K)$ & STL & Target Lasso \\
\midrule
3  & $(1.0,\,0.9,\,0.9)$
   & \textbf{0.2513} (0.0052) & 0.5475 (0.0074) \\
5  & $(1.0,\,0.9,\,0.9,\,0.9,\,0.0)$
   & \textbf{0.2628} (0.0050) & 0.5475 (0.0074) \\
10 & $(1.0,\,0.9,\,0.9,\,0.9,\,0.9,\,0.9,\,0.9,\,0.9,\,0.0,\,0.0)$
   & \textbf{0.2777} (0.0058) & 0.5475 (0.0074) \\
15 & $1.0,\ \text{then } 0.9 \text{ (}{\times}11\text{)},\ \text{then } 0.0 \text{ (}{\times}3\text{)}$
   & \textbf{0.2869} (0.0062) & 0.5475 (0.0074) \\
\bottomrule
\end{tabular}
\end{table}

In summary, the simulations validate the main theoretical predictions:
STL exploits informative external supports to reduce estimation
error below the Lasso rate, even when no support is individually
complete. The SVD-guided aggregation handles
collinearity and performs data-driven source selection, and the
residual-correction step provides a worst-case guarantee matching the
target-only Lasso.

\section{Real Data Analysis}
\label{sec:realdata}

DNA methylation adds small chemical tags, called methyl groups, to DNA. These tags can
attach at locations known as CpG positions. In a sample containing many cells, the
methylation level at a position is the proportion of DNA copies carrying the tag there.
At a small subset of CpG positions, this level increases or decreases steadily with age.
Models called epigenetic clocks use these age-related changes to predict a person's age,
and are now standard tools in aging research and in forensic age estimation
\citep{bocklandt2011epigenetic,horvath2013clock,hannum2013clock,horvathraj2018}. Each
clock considered here is a sparse linear model that predicts age from a weighted sum of
methylation levels: its coefficients are the weights, and its support is the small set of
CpG positions receiving nonzero weight. The support is itself of scientific interest, and
every published clock reports it.

We ask which CpG positions carry age information in the target study, and with what
weights. We consider the saliva DNA-methylation data of \citet{bocklandt2011epigenetic}
(Gene Expression Omnibus accession GSE28746), which contain $n = 66$ samples from
individuals aged $21$ to $55$ years. After the preprocessing distributed with the data,
$p = 27{,}578$ CpG positions remain, and the $0.12\%$ of missing values are imputed by
position-wise means. The response is chronological age, and $p \gg n$.
With $66$ samples and $27{,}578$ positions, identifying the relevant
positions from the target data alone is difficult, and external information can help.
Age-related methylation has been studied for more than a decade, and several clocks have
been published and validated, but data-based transfer is not applicable
here. Individual-level data behind the published clocks are often under controlled access,
and even when available, they come from other tissues, mostly blood, and often from a later
platform covering far more positions, so that most positions the clocks use are not measured
in the target study. Neither the data nor the fitted weights of the clocks can therefore be
pooled with the target sample or used as a shrinkage target. The positions that the clocks
select, however, remain meaningful: a position whose methylation tracks age in blood is a
candidate for tracking age in saliva.

Here the supports are provided by an LLM, which was given the identifiers of the measured
positions and a description of the target study. Through reasoning, the LLM
proposed $K=6$ supports, of sizes $7$, $12$, $5$, $8$, $12$ and $12$. It also cited the
published study behind each support: \citet{hannum2013clock}, \citet{koch2011signature},
\citet{weidner2014aging}, \citet{vidalbralo2016simplified}, \citet{levine2018phenoage}
and \citet{zhang2019improved}.
Each support consists of the positions that its source reports and that are also
measured in the target study.
The supports are fixed prior
to model fitting, and no outcome values of the target study were available to the LLM,
which was also instructed to exclude any source constructed using the target samples. The
sources differ from the target study in tissue, in measurement platform and in the
response quantity predicted, so the reliability of their supports is unknown, and which of
the six are informative cannot be determined in advance.

\begin{figure}[t]
    \centering
    \includegraphics[width=0.5\linewidth]{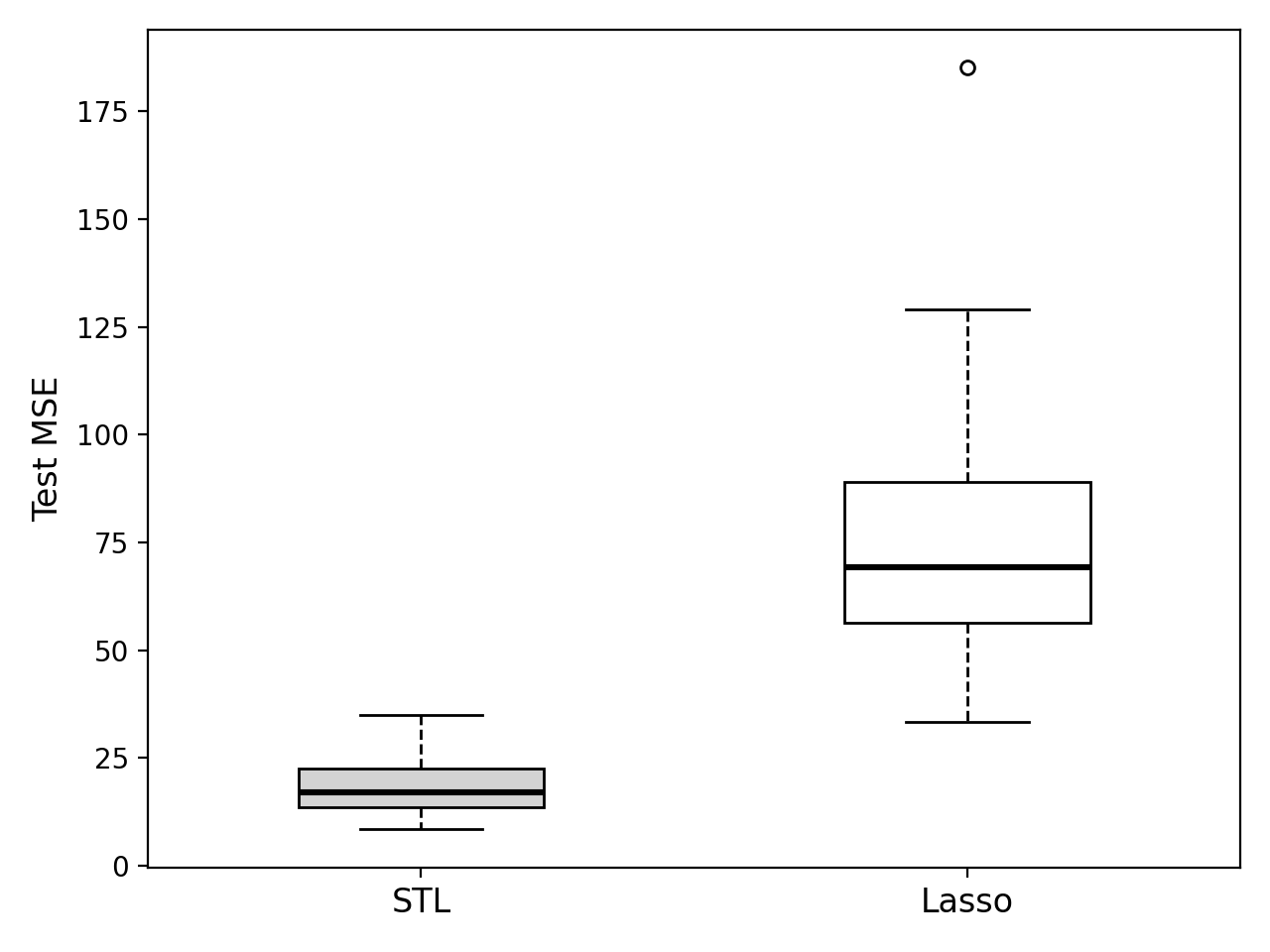}
    \caption{Test MSE over 30 random splits of the twin pairs for the proposed method and
the Lasso fitted to the target data alone}
    \label{fig:boxplot_mse}
\end{figure}

We compare the proposed method with the Lasso \citep{tibshirani1996lasso}
fitted to the target data alone and applied to all $27{,}578$ positions.
Performance is assessed by the mean squared error (MSE) of predicted age on held-out
samples. The $66$ samples come from $33$ twin pairs, so we split pairs rather than
individuals: we repeatedly assign $22$ pairs at random to training and the remaining $11$
to testing, fit each method on the $44$ training samples, and evaluate prediction on the
$22$ held-out samples, with the cross-validation and cross-fitting splits also keeping
twins together. Figure~\ref{fig:boxplot_mse} summarizes the distribution of test MSE over
$30$ such splits.
The proposed method attains a mean test MSE of $19.9$ years$^2$,
corresponding to a root mean squared error of $4.5$ years, against
$76.8$ years$^2$, or $8.8$ years, for the Lasso, and its error is the smaller of the two
in every one of the $30$ splits (paired Wilcoxon signed-rank test, $p$-value
$< 10^{-8}$). The Lasso selects $33$ positions on average, yet its mean test MSE is close
to the $80.7$ years$^2$ obtained by predicting the training mean age, which is consistent
with the well-documented instability of purely data-driven variable selection when
$n \ll p$. The fitted
aggregation of the proposed method is also interpretable: the sources of
\citet{hannum2013clock} and \citet{zhang2019improved} receive the largest coefficients,
while that of \citet{levine2018phenoage}, which targets mortality risk rather than
chronological age, receives a coefficient near zero. These coefficients are estimated
from the target data; the only external input is the set of positions each
source reports, and it roughly halves the root
mean squared error of predicted age.

\section{Discussion}

In this paper, we studied high-dimensional sparse learning with structured prior information in the form of externally suggested support sets, motivated in particular by supports proposed by large language models. The central challenge is that such supports can be highly heterogeneous and redundant, and they may be partially informative, noisy, or even entirely incorrect due to hallucinations, so naively selecting a single support or directly trusting external suggestions can lead to unstable estimation and biased inference. To address this, we proposed a robust transfer procedure that aggregates across all candidate supports by fitting restricted base models, learning data-adaptive combination weights in a cross-fitted manner, stabilizing the aggregation using an SVD-based penalization to handle strong collinearity among candidates, and finally applying a sparse residual correction to recover target-specific effects not captured by the external supports. The resulting method retains the target-only Lasso rate under an explicit limit on the number of spurious coordinates, attains improved rates under joint coverage and distinguishable source directions, and is minimax optimal under joint coverage with a possibly nonzero omitted signal, even when the supports contain hallucinated coordinates.
These guarantees clarify when LLM-generated structure can improve estimation and when residual correction controls its errors.

Practically, this framework enables more reliable use of LLMs as a source of structured hypotheses in data-limited scientific and industrial settings, supporting workflows that prioritize interpretability by focusing attention on coherent feature groups while remaining protected against spurious suggestions, and it can serve as a foundation for downstream tasks such as knowledge-guided modeling, robust decision support, and interpretable AI in domains where collecting new labels is expensive. Looking forward, several directions are promising, including extending the approach beyond linear models (for example to generalized linear models and survival outcomes), developing theory and algorithms for adaptive prompting or active acquisition of external supports, incorporating uncertainty quantification for the aggregated and corrected estimates, and studying robustness across prompt-induced support generators.

\bibliographystyle{chicago}
\bibliography{references}

\end{document}